\documentclass{emulateapj}
\usepackage{rotating}
\usepackage{url}

\newcommand{\msun}{\ifmmode M_{\odot}\else$M_{\odot}$\fi}
\newcommand{\rsun}{\ifmmode R_{\odot}\else$R_{\odot}$\fi}
\newcommand{\degrees}{\ifmmode^{\circ}\else$^{\circ}$\fi}
\newcommand{\amin}{\ifmmode^{\prime}\else$^{\prime}$\fi}
\newcommand{\asec}{\ifmmode^{\prime\prime}\else$^{\prime\prime}$\fi}
\newcommand{\fermi}{{\it Fermi}}
\newcommand{\gray}{$\gamma$-ray}
\newcommand{\grays}{$\gamma$-rays}
\newcommand{\mspa}{J0614$-$3329}
\newcommand{\mspb}{J1231$-$1411}
\newcommand{\mspc}{J2214$+$3000}

\shorttitle{MSPs in \fermi\ Sources}
\shortauthors{Ransom et al.}

\slugcomment{Accepted for publication in ApJ Letters}

\begin{document}

\title{Three Millisecond Pulsars in {\it FERMI}\ LAT Unassociated Bright Sources}

\author{
S.~M.~Ransom\altaffilmark{1,2},
P.~S.~Ray\altaffilmark{3,4},
F.~Camilo\altaffilmark{5},
M.~S.~E.~Roberts\altaffilmark{6},
\"O.~\c{C}elik\altaffilmark{7,8,9},
M.~T.~Wolff\altaffilmark{3},
C.~C.~Cheung\altaffilmark{10},
M.~Kerr\altaffilmark{11},
T.~Pennucci\altaffilmark{12},
M.~E.~DeCesar\altaffilmark{7,13},
I.~Cognard\altaffilmark{14},
A.~G.~Lyne\altaffilmark{15},
B.~W.~Stappers\altaffilmark{15},
P.~C.~C.~Freire\altaffilmark{16},
J.~E.~Grove\altaffilmark{3},
A.~A.~Abdo\altaffilmark{10},
G.~Desvignes\altaffilmark{17,18},
D.~Donato\altaffilmark{8,13},
E.~C.~Ferrara\altaffilmark{7},
N.~Gehrels\altaffilmark{7},
L.~Guillemot\altaffilmark{16},
C.~Gwon\altaffilmark{3},
A.~K.~Harding\altaffilmark{7},
S.~Johnston\altaffilmark{19},
M.~Keith\altaffilmark{19},
M.~Kramer\altaffilmark{15,16},
P.~F.~Michelson\altaffilmark{11},
D.~Parent\altaffilmark{20},
P.~M.~Saz~Parkinson\altaffilmark{21},
R.~W.~Romani\altaffilmark{11},
D.~A.~Smith\altaffilmark{22},
G.~Theureau\altaffilmark{14},
D.~J.~Thompson\altaffilmark{7},
P.~Weltevrede\altaffilmark{15},
K.~S.~Wood\altaffilmark{3},
M.~Ziegler\altaffilmark{21},
}
\altaffiltext{1}{National Radio Astronomy Observatory (NRAO), Charlottesville, VA 22903, USA}
\altaffiltext{2}{email: sransom@nrao.edu}
\altaffiltext{3}{Space Science Division, Naval Research Laboratory, Washington, DC 20375, USA}
\altaffiltext{4}{email: Paul.Ray@nrl.navy.mil}
\altaffiltext{5}{Columbia Astrophysics Laboratory, Columbia University, New York, NY 10027, USA}
\altaffiltext{6}{Eureka Scientific, Oakland, CA 94602, USA}
\altaffiltext{7}{NASA Goddard Space Flight Center, Greenbelt, MD 20771, USA}
\altaffiltext{8}{Center for Research and Exploration in Space Science and Technology (CRESST) and NASA Goddard Space Flight Center, Greenbelt, MD 20771, USA}
\altaffiltext{9}{Department of Physics and Center for Space Sciences and Technology, University of Maryland Baltimore County, Baltimore, MD 21250, USA}
\altaffiltext{10}{National Research Council Research Associate, National Academy of Sciences, Washington, DC 20001, resident at Naval Research Laboratory, Washington, DC 20375, USA}
\altaffiltext{11}{W. W. Hansen Experimental Physics Laboratory, Kavli Institute for Particle Astrophysics and Cosmology, Department of Physics and SLAC National Accelerator Laboratory, Stanford University, Stanford, CA 94305, USA}
\altaffiltext{12}{University of Virginia, Charlottesville, VA 22904, USA}
\altaffiltext{13}{Department of Physics and Department of Astronomy, University of Maryland, College Park, MD 20742, USA}
\altaffiltext{14}{ Laboratoire de Physique et Chimie de l'Environnement, LPCE UMR 6115 CNRS, F-45071 Orl\'eans Cedex 02, and Station de radioastronomie de Nan\c{c}ay, Observatoire de Paris, CNRS/INSU, F-18330 Nan\c{c}ay, France}
\altaffiltext{15}{Jodrell Bank Centre for Astrophysics, School of Physics and Astronomy, The University of Manchester, M13 9PL, UK}
\altaffiltext{16}{Max-Planck-Institut f\"ur Radioastronomie, Auf dem H\"ugel 69, 53121 Bonn, Germany}
\altaffiltext{17}{Department of Astronomy, University of California, Berkeley, CA 94720-3411, USA}
\altaffiltext{18}{Radio Astronomy Laboratory, University of California, Berkeley, CA 94720, USA}
\altaffiltext{19}{CSIRO Astronomy and Space Science, Australia Telescope National Facility, Epping NSW 1710, Australia}
\altaffiltext{20}{College of Science, George Mason University, Fairfax, VA 22030, resident at Naval Research Laboratory, Washington, DC 20375, USA}
\altaffiltext{21}{Santa Cruz Institute for Particle Physics, Department of Physics and Department of Astronomy and Astrophysics, University of California at Santa Cruz, Santa Cruz, CA 95064, USA}
\altaffiltext{22}{Universit\'e Bordeaux 1, CNRS/IN2p3, Centre d'\'Etudes Nucl\'eaires de Bordeaux Gradignan, 33175 Gradignan, France}

\begin{abstract}
  We searched for radio pulsars in 25 of the non-variable,
  unassociated sources in the \fermi\ LAT Bright Source List with the
  Green Bank Telescope at 820\,MHz.  We report the discovery of three
  radio and \gray\ millisecond pulsars (MSPs) from a high Galactic
  latitude subset of these sources.  All of the pulsars are in binary
  systems, which would have made them virtually impossible to detect
  in blind \gray\ pulsation searches.  They seem to be relatively
  normal, nearby ($\le$2\,kpc) millisecond pulsars.  These
  observations, in combination with the \fermi\ detection of \grays\
  from other known radio MSPs, imply that most, if not all, radio MSPs
  are efficient \gray\ producers.  The \gray\ spectra of the pulsars
  are power-law in nature with exponential cutoffs at a few GeV, as
  has been found with most other pulsars.  The MSPs have all been
  detected as X-ray point sources.  Their soft X-ray luminosities of
  $\sim$10$^{30-31}$\,erg\,s$^{-1}$ are typical of the rare radio MSPs
  seen in X-rays.
\end{abstract}


\section{Introduction}

Before the launch of the \fermi\ Gamma-ray Space Telescope, the only
pulsars with definitive detections in \grays\ (using EGRET on {\it
  CGRO}) were young and very energetic ($\dot
E>10^{36}$\,erg\,s$^{-1}$) or nearby older systems ($\dot
E>10^{34}$\,erg\,s$^{-1}$) \citep{thompson04}.  A possible detection
of pulsed \grays\ from the energetic MSP J0218$+$4232
\citep{egret0218}, sparked interest in modeling MSP \gray\ emission
\citep[e.g.][]{zc03,hum05}, and encouraged one group \citep{sgh07} to
predict that many new MSPs might be detected in \grays\ or discovered
in radio follow-up of unidentified \fermi\ sources.

The launch of \fermi\ and the extraordinary sensitivity of the Large
Area Telescope \citep[LAT,][]{atwood09}, confirmed those predictions
of \gray-bright MSPs with detections of eight relatively normal radio
MSPs using only the first few months of \fermi\ events
\citep{fermimsps}.  Those MSPs were detected via the folding of
\grays\ modulo the known spin and orbital ephemerides from radio
timing campaigns \citep{fermiptc}.

In order to best utilize radio telescope time to search either for
radio counterparts to new \gray-selected pulsars or to search blindly
for radio pulsations from \gray\ sources that might contain pulsars,
we formed the Pulsar Search Consortium (PSC), a group of approximately
20 LAT-team members and/or pulsar experts associated with large radio
telescopes around the world.  This paper describes one of the PSC's
first programs, which used the Green Bank Telescope (GBT) to search 25
unassociated sources from the \fermi\ LAT Bright Source List
\citep{fermibsl}.

\section{Observations and Data Analysis}

We selected 25 sources from the \fermi\ LAT Bright Source List that
were a) unassociated with known pulsars or active galactic nuclei
(AGN) b) unassociated with X-ray counterparts that had been previously
deeply searched for radio pulsations (e.g. IC443, Camilo et al., in
prep.) c) statistically non-variable and d) at declinations
$>$$-$35\degrees.  We observed each of the sources\footnote{The LAT
  Bright Sources observed with the GBT were 0FGLs J0614.3$-$3330,
  J1231.5$-$1410, J1311.9$-$3419, J1653.4$-$0200, J1741.4$-$3046,
  J1746.0$-$2900, J1801.6$-$2327, J1805.3$-$2138, J1814.3$-$1739,
  J1821.4$-$1444, J1834.4$-$0841, J1836.1$-$0727, J1839.0$-$0549,
  J1844.1$-$0335, J1848.6$-$0138, J1855.9$+$0126, J1900.0$+$0356,
  J1911.0$+$0905, J1923.0$+$1411, J2001.0$+$4352, J2027.5$+$3334,
  J2110.8$+$4608, J2214.8$+$3002, J2302.9$+$4443, and J2339.8$-$0530.}
for approximately 45$-$50\,minutes using the prime focus receiver at
the GBT centered at 820\,MHz with 200\,MHz of bandwidth.  The GBT
pointings, all taken between July and October 2009, were actually
centered on the positions from an internal LAT source list using nine
months of sky-survey data prepared in a similar fashion to the \fermi\
LAT First Source Catalog \citep[i.e. ``1FGL'',][]{1FGLcat}.  The
individual GBT pointings at 820\,MHz had FWHM$=$0.25\degrees\ and
covered either all or a substantial fraction of the 95\% error regions
for the vast majority of the sources.

We sampled the summed power from two polarizations in 2048 frequency
channels with 8-bits every 61.44\,$\mu$s using the GUPPI pulsar
backend\footnote{\url{https://safe.nrao.edu/wiki/bin/view/CICADA/NGNPP}}.
Each pointing generated approximately 100\,GB of data, which were
recorded to hard drives for processing off-site.  The 820\,MHz center
frequency was chosen as a compromise between the competing effects of
sky temperature (from the Galactic synchrotron background) and
beamsize, as well as steep pulsar spectra and the effects of
interstellar dispersion and scattering.

For a pulsar with a pulse width of $\sim$10\% of the period, the
search sensitivity was approximately $0.06\times(29\,{\rm K} + T_{\rm
  sky})/(32\,{\rm K})$\,mJy, where $T_{\rm sky}$ is the contribution
at 820\,MHz of the Galactic synchrotron background.  The majority of
the sources (17 of them) were within three degrees of the Galactic
plane where $T_{\rm sky}\sim12-40$\,K, although two were very near the
Galactic center with $T_{\rm sky}\sim 100-150$\,K.  Eight of the
sources were well off the Galactic plane ($|b|>5$\degrees) and had
$T_{\rm sky}\sim 3-10$\,K.  For those sources, our search sensitivity
was 0.06$-$0.08\,mJy for normal pulsars at all reasonable dispersion
measures (DMs) and MSPs up to DM$\sim$100\,pc\,cm$^{-3}$.  In general,
the observations were factors of 2$-$12 deeper than the best pulsar
surveys that have previously covered these regions
\citep[e.g.][]{mlc+01,cfl+06}.

We processed the data, after de-dispersing into $\sim$9000 DMs over
the range 0$-$1055\,pc\,cm$^{-3}$, using both acceleration searches
(to improve sensitivity to pulsars in binary systems) and single pulse
searches (to provide sensitivity to pulsars with sporadic or
giant-pulse-like emission) using standard tools found in
{\textsc{PRESTO}}\footnote{\url{http://www.cv.nrao.edu/~sransom/presto/}}\citep{rem02}.
No new pulsar-like signals were found in any of the low Galactic
latitude ($|b|<5$\degrees) sources.  However, four new MSPs were
detected amid the eight high Galactic latitude sources, in 0FGLs\
J0614.3$-$3330, J1231.5$-$1410, J2214.8$+$3002, and J2302.9$+$4443.
The MSP in 0FGL J2302.9$+$4443 was detected first in an independent
PSC survey by the Nan\c{c}ay telescope and will be reported elsewhere
(Cognard et al.~in prep.).  The rest of this paper details the
properties of the other three MSPs.

\subsection{The New MSPs}

The first two pulsars detected, \mspc\ and \mspb, were undergoing
substantial accelerations due to orbital motion during the discovery
observations.  PSR \mspa\ was initially uncovered in an unaccelerated
search, although orbital motion was detected in the discovery
observations via a more precise timing analysis.  The fact that all
three MSPs were in unknown binaries demanded a radio timing program to
determine precise orbital parameters and constrain their astrometric
positions before detailed \gray\ timing and analysis could commence.

Each MSP was observed with several different observing setups at the
GBT, the Lovell Telescope at Jodrell Bank, and the Nan\c{c}ay radio
telescope.  The Arecibo telescope also observed PSR \mspc\ several
times.  At the GBT, GUPPI was used with bandwidths of 100, 200\, and
800\,MHz centered at 350, 820, and 1500\,MHz respectively.  At Jodrell
Bank and Nan\c{c}ay, observations were made with bandwidths of
200$-$300\,MHz centered near 1400\,MHz.  Standard radio timing
procedures were used (Lorimer \& Kramer 2005)\nocite{lk05} and the
orbital parameters were fit to high precision with {\tt TEMPO2}
\citep{hem06}.

After several months of radio timing, and using trial X-ray positions
based on point sources from \textit{Swift}, \textit{XMM-Newton},
and/or \textit{Chandra}, \gray\ pulsations from each MSP were detected
with the LAT using orbitally-demodulated events (see also
\S\ref{gray}).  We determined average \gray\ pulse Times of Arrival
(TOAs) using the maximum likelihood \gray\ timing techniques described
by \citet{ray10} after integrating source photons modulo the predicted
pulse period for between 22 and 36 days per TOA.  The resulting joint
timing solutions, using both radio and \gray\ TOAs, as well as the
derived physical parameters of the MSPs, are presented in Table~1.
The radio and \gray\ pulse profiles are shown in
Figure~\ref{Fig:Seds}.

The three new pulsars, besides being three of the brightest \gray\
MSPs in the sky, appear to be relatively normal, nearby ($\le$2\,kpc),
radio MSPs, with $\sim$3\,ms spin periods, surface magnetic field
strengths of (2$-$3)$\times$10$^8$\,G, and spin-down luminosities of
$\sim$2$\times$10$^{34}$\,erg\,s$^{-1}$.  \mspa\ and \mspb\ have
orbital periods of 53.6 and 1.9\,days respectively, with companions of
mass $\sim$0.2$-$0.3\,\msun, consistent with the orbital period -
white dwarf mass relation of \citet{rpj+95}.  \mspc\ is a so-called
``black-widow'' system with a very low-mass companion
($\sim$0.02\,\msun) and likely timing irregularities, similar to
pulsars B1957$+$20 \citep{fst88}, J2051$-$0827 \citep{sbl+96}, and
J0610$-$2100 \citep{bjd+06}, and only the fourth such system known in
the Galactic disk.  While we currently have no evidence for radio
eclipses from the pulsar (at least at frequencies $\ge$1.4\,GHz), its
formation was likely similar to that of those other systems
\citep[e.g.][]{kbr+05}.

\subsection{\gray\ Analysis}
\label{gray}

The \fermi\ LAT is sensitive to \grays\ with energies
0.02$-$300\,GeV~\citep{atwood09}.  The \fermi\ LAT sky-survey data set
used here for spectral analysis spans from 2008 August 4 to 2010
February 4.  We selected ``Pass 6 Diffuse'' class events -- i.e.
events passing the most stringent background rejection cuts -- with
energies above 0.1\,GeV and rejected events with zenith angles
$>105\degrees$ to limit contamination from \grays\ from the Earth's
limb.  We used ``Pass6 v3'' instrument response functions (IRFs).

\begin{figure}[tbhp]
 \centering
 \includegraphics[width=\columnwidth,angle=270]{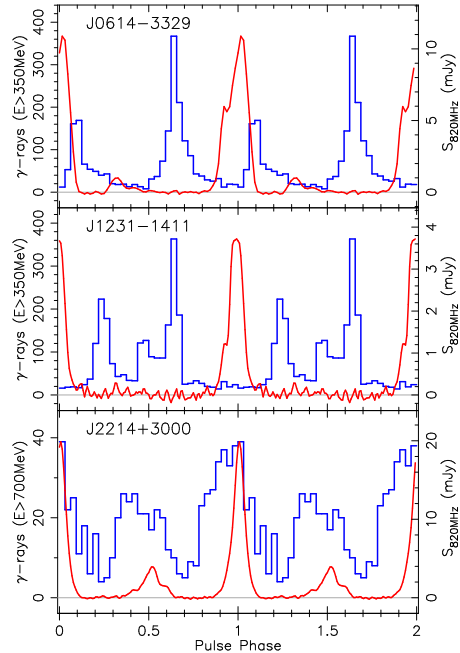}
 \caption{Radio and \gray\ pulse profiles for the three new MSPs. The
   red lines correspond to the 820\,MHz discovery pulse profiles from
   the GBT with the 820\,MHz flux density scale on the right.  The
   blue lines are the {\it Fermi}\ pulse profiles with the photon
   counts and low energy cut used for the \gray\ selections listed on
   the left.  There are 2043, 2341, and 621 photons in the \gray\
   profiles for PSRs \mspa, \mspb, and \mspc, respectively.}
 \label{Fig:Profs}
\end{figure}

The \gray\ light curves shown in Figure~\ref{Fig:Profs} are
constructed from events as described above, although using seven
additional months of data through 2010 September 14, and with energy
and radius cuts to optimizes the signal to noise for each pulsar.  The
energy and radius cuts used for PSRs \mspa, \mspb, and \mspc\ are
(1.0$^\circ$, 0.35\,GeV), (1.2$^\circ$, 0.35\,GeV), and (1.0$^\circ$,
0.7\,GeV), respectively.

We derived the \gray\ spectrum of each pulsar using a
maximum-likelihood method implemented in the LAT {\it Science Tool}
{\tt
  gtlike}\footnote{\url{http://fermi.gsfc.nasa.gov/ssc/data/analysis/documentation/}}.
We analyzed a region of 10\degrees\ radius centered on the radio
position of each pulsar, and modeled each region by including all
sources from the 1FGL~\citep{1FGLcat} within 17\degrees\ of the pulsar
along with Galactic and isotropic diffuse emission (models
\texttt{gll\_iem\_v02} and \texttt{isotropic\_iem\_v02},
respectively\footnote{\url{http://fermi.gsfc.nasa.gov/ssc/data/access/lat/BackgroundModels.html}}).
The power-law spectral parameters for all sources within 10\degrees\
of the pulsar and a normalizing scale factor for the diffuse emission
spectrum were allowed to be free in the fit.

\begin{figure}[tbhp]
 \centering
 \includegraphics[width=\columnwidth]{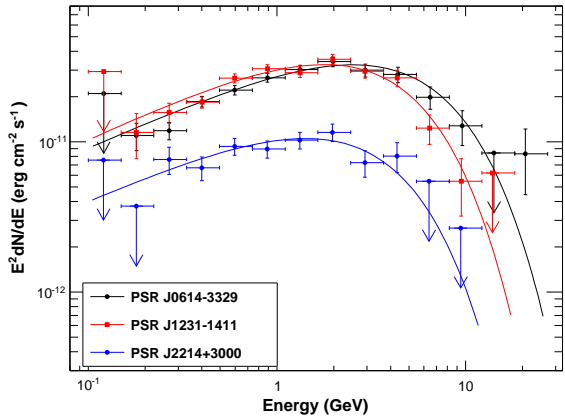}
 \caption{The \gray\ spectra for the three MSPs. The flux points on
   the curve were obtained from independent fits in each energy bin,
   as explained in the text. The curves represent the spectrum fit
   with a simple exponentially cut off power-law in the full energy
   range of $0.1-100$\,GeV.}
 \label{Fig:Seds}
\end{figure}

We modeled the spectrum of each MSP using a power-law with an
exponential cutoff where the three parameters, the differential flux
$K$, the photon index $\Gamma$, and the cutoff energy $E_{\rm
  cutoff}$, were allowed to vary in the fit. The phase-averaged \gray\
spectra obtained for each pulsar are shown in Figure~\ref{Fig:Seds}
and the spectral parameters are given in Table~1.  The uncertainty in
the LAT effective area is estimated to be $\le$5\% near 1\,GeV, 10\%
below 0.1\,GeV and 20\% over 10\,GeV. The resulting systematic errors
on the three spectral parameters, propagated from the uncertainties on
the LAT effective area, were calculated using a set of ``modified
IRFs'' bracketing the nominal (Pass6 v3) one.

We verified the significance of the exponential cutoff in each
spectrum with a likelihood ratio test \citep{mbc+96}.  A simple
power-law model is rejected significantly for all three pulsars
relative to an exponentially cutoff power-law, as indicated by the
test statistic $\Delta$TS$_{\rm cutoff}$ listed in Table~1 for the
addition of one free parameter.

The flux points in Figure~\ref{Fig:Seds} were obtained by repeating
the likelihood analysis in each energy band, assuming a power-law
spectrum with a photon index fixed at 2 and a free flux normalization
parameter for all sources.

\subsection{X-ray Analysis}

To search for X-ray counterparts, we observed the field of each pulsar
with the \textit{Swift}-XRT \citep{sxrt} with exposures ranging from
2.6$-$15.9\,ks. For \mspa\ and \mspb\ we detected X-ray point sources
at the locations of the pulsars. In the Swift observation of PSR
\mspc, there is no significant source at the location of the pulsar,
however we also obtained a {\it Chandra} ACIS-I observation of this
region and detected the pulsar with it.  A detailed analysis of this
{\it Chandra} observation will be presented elsewhere.

The X-ray counterparts to the MSPs are soft sources and were fit to
black body spectra using XSPEC, fixing the absorption values to zero.
The resultant temperatures for the three X-ray sources were in the
range $\sim$0.21$-$0.25 keV, which in the case of \mspb, is consistent
with that derived from deeper Suzaku data (Maeda et al.~2010, in
prep.).  Allowing for additional Galactic absorption fixed to the
values of $(3.4-5.7)\times 10^{20}$\,cm$^{-2}$ from \citep{kbh+05}, we
found negligible differences in the fitted parameters.  The
\textit{Swift} and \textit{Chandra} positional localizations, as well
as approximate fluxes and luminosities from the black body fits, are
in Table~1.

For PSR \mspb\ we were able to do more detailed spectral analysis
using \textit{XMM-Newton}. On 2009 July 15 \textit{XMM-Newton}
observed the field of this as-yet unidentified BSL source with all
three EPIC instruments.  Data from each instrument were analyzed
utilizing the Science Analysis System software version 10.0.0 and the
calibration update of May 8, 2010. We filtered the data for bad events
and excluded times of high particle background, yielding 29.4\,ks and
29.5\,ks of good time for the EPIC-MOS1 and -MOS2 instruments and
24.1\,ks for EPIC-PN.  All three instruments utilized Full Frame mode
so none had sufficient time resolution to allow searches for X-ray
pulsations.

The \textit{XMM-Newton} X-ray images of the \mspb\ field reveal a
relatively isolated, moderately bright, point source that we name
XMMUJ123112$-$141146 at the best radio position of the pulsar to the
accuracy of the EPIC instruments. We generated spectra of this source
by extracting events from around the pulsar in 35\asec\ regions for
the MOS images and a 25\asec\ region (due to a chip gap) for the PN
image. This yielded spectra consisting of 853, 812, and 2164 events
from the MOS1, MOS2, and PN cameras, respectively. We group the counts
with at least 30 events per spectral bin for the MOS spectra and 45
events per bin for the PN spectra using the FTOOL {\tt grppha}.
Background spectra were extracted from nearby regions 100\asec\ and
55\asec\ in radius from the same CCD chips for the MOS and PN
instruments, respectively.

Using XSPEC (v12.6.0) we simultaneously fit the MOS1, MOS2, and PN
spectra in the energy range 0.4$-$3.0 keV. An absorbed power-law model
is formally acceptable with a reduced $\chi^2$=0.977 (61 dof) but with
an extremely steep photon index of $\Gamma =4.23^{+0.41}_{-0.38}$
(90\% confidence) and column density of $n_H=(1.8^{+0.6}_{-0.5})\times
10^{21}$\,cm$^{-2}$ (90\% confidence).  Such a column density is
significantly higher than that expected for this direction based on
galactic surveys \citep[$n_H=3.45\times
10^{20}$\,cm$^{-2}$,][]{kbh+05}. A fit to these data utilizing a model
of an absorbed neutron star non-magnetic hydrogen atmosphere
\citep[{\em phabs$\times$nsatmos},][]{hrn+06} with the neutron star
mass and radius held fixed at 1.4\msun\ and 10\,km, and the source
distance fixed at the dispersion measure value of 0.4\,kpc, yields a
reduced $\chi^2$=1.41 (61 dof). In this model there are significant
residuals above 1.5\,keV so we add a power-law component with photon
index fixed at 1.8 and obtain an improved fit with $\chi^2$=1.09 (60
dof). This latter fit yields an atmospheric temperature (seen at
infinity) $T_\mathrm{eff}=61^{+6}_{-12}$\,eV (90\% confidence), a best
fit $n_H$ consistent with zero and a 90\% confidence upper limit of
$5\times 10^{20}$\,cm$^{-2}$, and flux in the 0.5$-$3\,keV energy band
$(1.15\pm0.05)\times 10^{-13}$\,ergs\,cm$^{-2}$\,s$^{-1}$.  In this
model the power-law spectral component accounts for roughly 25\% of
the total flux in the 0.5$-$3\,keV energy band.

\section{Conclusions}

We have identified three new nearby radio MSPs as the counterparts of
bright and previously unassociated {\it Fermi} LAT sources at high
Galactic latitude.  Our non-detection of young pulsars or MSPs in the
more numerous sources searched at low Galactic latitude is likely due
to our only moderate sensitivity improvements (typically
2$-$3$\times$) over the best surveys of those regions to date
\citep[e.g.][]{mlc+01} due to higher sky temperatures resulting from
our lower observing frequency.  Additionally, the complicated and
confused nature of the Galactic plane in \grays\ makes the positive
identification of point sources difficult. Several of the bright
sources may be blends of other sources or the result of insufficient
modelling of the Galactic background.  Nonetheless, deeper surveys at
frequencies of 1.5$-$2\,GHz of these sources may prove more fruitful
in the future.

The new pulsars are very typical radio MSPs in terms of spin period,
binary parameters, magnetic field strength, spin-down luminosity, and
characteristic age, and their unusual brightness in \grays\ is likely
due more to their proximity than to especially energetic emission
processes in their magnetospheres.  The very high implied \gray\
efficiency for PSR \mspa\ suggests it is likely closer, by up to a
factor of 2 or more, than predicted by the NE2001 model \citep{cl02}.
The line-of-sight to PSR \mspa\ is nearly tangent to the Gum Nebula
where NE2001 shows an exceptionally steep DM gradient.  Additionally,
the pulsar's \gray\ emission is likely not isotropic, but only covers
tens of percent of the sky.  These large efficiencies in general,
though, are consistent with the tens of percent values found by
\citet{fermimsps} for radio MSPs detected in \grays\ and imply that
MSPs are very efficient producers of \grays.

We do not have proper motion measurements for pulsars \mspa\ or \mspb\
and so their measured spin-down rates are contaminated at some level
(likely $\lesssim$10\%) by the Shklovskii effect \citep{shk70}.  There
is a statistically significant proper motion measurement of
$\sim$100\,mas\,yr$^{-1}$ for \mspb, though, which implies a
Shklovskii effect at 400\,pc larger than the measured spin-down rate
for the pulsar.  If the proper motion is confirmed at this level, the
requirement to have the pulsar intrinsically spinning down gives an
upper limit for the pulsar's distance of $\sim$240\,pc.  Timing
observations over the next several years will determine the proper
motions and possibly the timing parallaxes for each of the pulsars.

In all three cases we identified X-ray counterparts to the pulsars
which substantially aided in the rapid establishment of timing
solutions.  The three MSPs appear to have fairly typical X-ray
properties for radio MSPs \citep[e.g.][]{bgh+06} with primarily soft
thermal-like spectra and X-ray luminosities in the
10$^{30-31}$\,erg\,s$^{-1}$ range, approximately 10$^{-4}$ to
10$^{-3}$ of their \gray\ luminosities.

The radio flux densities of $\sim$1\,mJy near 1\,GHz are large enough
to make the MSPs potentially useful for a wide variety of timing
projects, such as the detection of gravitational waves via long-term
pulsar timing (e.g. NANOGrav\footnote{\url{http://nanograv.org}}), yet
they are small enough to explain why earlier large-area surveys for
pulsars missed them \citep[e.g.][]{mld+96,lxf+05}.  In addition, the
fact that many of the nearby radio MSPs are being detected in \grays\
and vice-versa argues that the sizes of the radio and \gray\ beams are
comparable for MSPs (likely within a factor of $\sim$2), and that deep
radio and \gray\ surveys may allow us to eventually detect a large
percentage of the local population of these sources.  In the
short-term, the fact that {\it Fermi} can point us to nearby radio
MSPs is already causing a large increase in the number of known
systems, with much less effort than is required by sensitive
large-area radio surveys.  If most radio MSPs produce \grays\ as these
early results seem to indicate, MSPs may contribute to the diffuse
isotropic \gray\ background \citep{fl10}.

\acknowledgments We acknowledge helpful discussions with Natalie Webb
and Lynne Valencic. The National Radio Astronomy Observatory is a
facility of the National Science Foundation operated under cooperative
agreement by Associated Universities, Inc.  This work was partially
supported by NASA Grant No.\ NNG09EE57I.  The \textit{Fermi} LAT
Collaboration acknowledges support from a number of agencies and
institutes for both development and the operation of the LAT as well
as scientific data analysis.  These include NASA and DOE in the US,
CEA/Irfu and IN2P3/CNRS in France, ASI and INFN in Italy, MEXT, KEK,
and JAXA in Japan, and the K.~A.~Wallenberg Foundation, the Swedish
Research Council and the National Space Board in Sweden.  Additional
support from INAF in Italy and CNES in France for science analysis
during the operations phase is also gratefully acknowledged.

{\it Facilities:} \facility{GBT (GUPPI)}, \facility{Fermi (LAT)},
\facility{XMM (EPIC)}, \facility{CXO (ACIS)}, \facility{Swift (XRT)}


\begin{deluxetable}{lccc}
\tabletypesize{\footnotesize}
\tablecaption{Parameters for the New MSPs\label{tab1}}
\tablewidth{0pt}
\tablehead{\colhead{Parameter} & \colhead{PSR J0614$-$3329} & \colhead{PSR J1231$-$1411} & \colhead{PSR J2214$+$3000}}
\startdata
{\it Fermi}\ BSL Association (0FGL)\dotfill & J0614.3$-$3330 & J1231.5$-$1410 & J2214.8$+$3002 \\
{\it Fermi}\ 1-Year Source (1FGL)\dotfill & J0614.1$-$3328 & J1231.1$-$1410 & J2214.8$+$3002 \\
\cutinhead{Timing Parameters}
Right Ascension (RA, J2000) \dotfill & $06^{\rm h}\;14^{\rm m}\;10\fs3478(3)$ & $12^{\rm h}\;31^{\rm m}\;11\fs3132(7)$ & $22^{\rm h}\;14^{\rm m}\;38\fs8460(1)$ \\
Declination     (DEC, J2000) \dotfill & $-33\degrees\;29\amin\;54\farcs118(4)$ & $-14\degrees\;11\amin\;43\farcs63(2)$ & $+30\degrees\;00\amin\;38\farcs234(4)$ \\
Proper Motion in RA (mas\,yr$^{-1}$) \dotfill & \dots & -1.0(2)$\times$10$^{2}$ & \dots \\
Proper Motion in DEC (mas\,yr$^{-1}$) \dotfill & \dots & -3(4)$\times$10$^{1}$ & \dots \\
Pulsar Period (ms) \dotfill & 3.148669579439(9) & 3.683878711077(3) & 3.119226579079(4) \\
Pulsar Frequency (Hz) \dotfill & 317.5944552995(9) & 271.4530196103(2) & 320.5922925597(4) \\
Frequency Derivative (Hz\,s$^{-1}$) \dotfill & -1.77(7)$\times$10$^{-15}$ & -1.68(1)$\times$10$^{-15}$ & -1.44(3)$\times$10$^{-15}$ \\
Frequency 2nd Deriv. (Hz\,s$^{-2}$) \dotfill & \dots & \dots & 1.7(4)$\times$10$^{-23}$ \\
Reference Epoch (MJD) \dotfill & 55100 & 55100 & 55100 \\
Dispersion Measure (pc cm$^{-3}$) \dotfill & 37.049(1) & 8.090(1) & 22.557(1) \\
Orbital Period (days) \dotfill & 53.5846127(8) & 1.860143882(9) & 0.416632943(5) \\
Projected Semi-Major Axis (lt-s)  \dotfill & 27.638787(2) & 2.042633(3) & 0.0590800(9) \\
Orbital Eccentricity  \dotfill & 0.0001801(1) & 4(3)$\times$10$^{-6}$ & $<2\times10^{-4}$ \\
Longitude of Periastron (deg) \dotfill & 15.92(4) & 3.2(4)$\times$10$^{2}$ & \dots \\
Epoch of Periastron (MJD) \dotfill & 55146.821(7) & 55016.8(2) & \dots \\
Epoch of Ascending Node (MJD) \dotfill & \dots & \dots & 55094.137854(2) \\
Span of Timing Data (MJD) \dotfill & 54683$-$55422 & 54683$-$55430 & 54683$-$55415 \\
Number of $\gamma$-ray TOAs \dotfill & 24 & 32 & 20 \\
RMS $\gamma$-ray TOA Residual ($\mu$s) \dotfill & 99.1 & 24.4 & 110.2 \\
Number of radio TOAs  \dotfill & 328 & 136 & 437 \\
RMS radio TOA Residual ($\mu$s) \dotfill & 7.1 & 9.3 & 5.0 \\
\cutinhead{Derived Parameters}
Mass Function (\msun) \dotfill & 0.007895133(3) & 0.00264460(2) & 1.2755(1)$\times$10$^{-6}$ \\
Min Companion Mass (\msun) \dotfill & $\geq$\,0.28 & $\geq$\,0.19 & $\geq$\,0.014 \\
Galactic Longitude (deg) \dotfill & 240.50 & 295.53 & 86.86 \\
Galactic Latitude (deg) \dotfill & -21.83 & 48.39 & -21.67 \\
DM-derived Distance (kpc) \dotfill & 1.9 & 0.4 & 1.5 \\
Flux Density at 820\,MHz (mJy) \dotfill & 1.5 & 0.4 & 2.1 \\
Surface Magnetic Field ($10^8$\,G) \dotfill & 2.4 & 2.9 & 2.1 \\
Characteristic Age (Gyr) \dotfill & 2.8 & 2.6 & 3.5 \\
Spin-down Lumin, $\dot E$ ($10^{34}$\,ergs\,s$^{-1}$) \dotfill & 2.2 & 1.8 & 1.8 \\
\cutinhead{\gray\ Spectral Fit Parameters}
K ($10^{-11}$ ph\,cm$^{-2}$\,s$^{-1}$\,MeV$^{-1}$) \dotfill  & 2.12 $\pm$ 0.10 $\pm$ 0.13 & 2.62 $\pm$ 0.16 $\pm$ 0.18 & 0.94 $\pm$ 0.11 $\pm$ 0.05\\
Spectral Index $\Gamma$ \dotfill  & 1.44 $\pm$ 0.05 $\pm$ 0.07 & 1.40 $\pm$ 0.07 $\pm$ 0.05 & 1.44 $\pm$ 0.13 $\pm$ 0.11\\
E$_{\rm cutoff}$ (GeV)  \dotfill  & 4.49 $\pm$ 0.54 $^{+1.38}_{-0.84}$ & 2.98 $\pm$ 0.33 $^{+0.43}_{-0.29}$ & 2.53 $\pm$ 0.50 $^{+0.43}_{-0.29}$\\
F$_{100}$ ($10^{-8}$ ph\,cm$^{-2}$\,s$^{-1}$) \dotfill  & 9.52 $\pm$ 0.46 $\pm$ 0.45 & 10.57 $\pm$ 0.62 $\pm$ 0.39 & 3.83 $\pm$ 0.44 $\pm$ 0.06\\
G$_{100}$ ($10^{-11}$ ergs\,cm$^{-2}$\,s$^{-1}$) \dotfill  & 10.86 $\pm$ 0.35 $\pm$ 1.06 & 10.33 $\pm$ 0.35 $\pm$ 0.87 & 3.32 $\pm$ 0.21 $\pm$ 0.24\\
TS \dotfill & 5270.3 & 4798.4 & 958.0 \\
$\Delta$TS$_{\rm cutoff}$ \dotfill & 184.7 & 203.3 & 63.3 \\
$\eta$ (\%) \dotfill & 210 & 11 & 49 \\
\cutinhead{X-ray Parameters}
X-ray Source RA (J2000) \dotfill & $06^{\rm h}\;14^{\rm m}\;10\fs3(3)$ & $12^{\rm h}\;31^{\rm m}\;11\fs3(4)$ & $22^{\rm h}\;14^{\rm m}\;38\fs84(3)$ \\
X-ray Source Dec (J2000)\dotfill & $-33\degrees\;29\amin\;54(5)$ & $-14\degrees\;11\amin\;43(6)$ & $30\degrees\;00\amin\;38\farcs2(6)$ \\
BB Temperature (keV) \dotfill & 0.23(5) & 0.21(5) & 0.25(4) \\
$F_{{\rm BB},0.5-8\,{\rm keV}}$ ($10^{-14}$ ergs\,cm$^{-2}$\,s$^{-1}$) \dotfill & 8.7 $^{+3.4}_{-3.9}$ & 15 $^{+5.3}_{-7.4}$ & 2.9 $^{+0.6}_{-0.7}$ \\
$L_{{\rm BB},0.5-8\,{\rm keV}}$ ($10^{30}$ ergs\,s$^{-1}$) \dotfill & 38 $^{+15}_{-17}$ & 2.9 $^{+1.0}_{-1.4}$ & 7.8 $^{+1.6}_{-1.9}$
\enddata

\tablecomments{Numbers in parentheses represent 2-$\sigma$
  uncertainties in the last digit as determined by {\tt TEMPO2} using
  the DE405 Solar System Ephemeris for the timing parameters and
  1-$\sigma$ uncertainties for the other parameters.  The time system
  used is Barycentric Dynamical Time (TDB). Minimum companion masses
  were calculated assuming a pulsar mass of 1.4\,\msun. The
  DM-distances were estimated using the NE2001 Galactic electron
  density model and likely have $\sim$20\% uncertainties \citep{cl02}.
  The gamma-ray spectral parameters are from fits of exponentially
  cutoff power-laws as described in \S2.2. F$_{100}$ and G$_{100}$
  give the integrated photon or energy flux above 0.1\,GeV,
  respectively, while the last two parameters are gamma-ray detection
  significance of the source and significance of an exponential cutoff
  (as compared to a simple power-law), where the approximate Gaussian
  significance is given by $\sim\sqrt{\rm TS}$, and TS is the Test
  Statistic TS = $2\Delta\log({\rm likelihood})$ between models with
  and without the source. The first errors are statistical and the
  second errors are systematic errors calculated from the bracketing
  IRFs.  The $\gamma$-ray efficiency, $\eta =L_\gamma f_\Omega/\dot
  E=4\pi D^2G_{100}/\dot E$, assumes a beaming correction factor,
  $f_\Omega=1$.  The X-ray results are from {\it Swift} for PSRs
  J0614$-$3329 and J1231$-$1411, and {\it Chandra} for PSR
  J2214$+$3000.  The quoted positional uncertainty for J2214$+$3000 is
  dominated by a systematic error of 0.6\asec\ in the absolute {\it
    Chandra} astrometry (statistical error $=$0.2\asec).  X-ray fluxes
  and luminosities from 0.5$-$8\,keV, $F_{{\rm BB},0.5-8\,{\rm keV}}$
  and $L_{{\rm BB},0.5-8\,{\rm keV}}$, are based on black body fits to
  the point source counts using $kT=0.16$\,keV.}

\end{deluxetable}

\end{document}